\documentclass[a4paper,11pt]{article}
\usepackage[margin=1in]{geometry}

\usepackage{microtype}

\usepackage{hyphenat}

\usepackage[capitalize]{cleveref}
\usepackage{booktabs}  %% tables
\usepackage{comment}
\usepackage[inline]{enumitem}
\usepackage{authblk} % Improved author formatting

\usepackage{navigator}
\usepackage{url}

\usepackage{float}

\usepackage{pdfpages}

\usepackage{todonotes}

\title{\centering Security at the Border? \\The Lived Experiences of Refugees and Asylum Seekers in the UK}

\author{Arshia Dutta\\ \texttt{arshia.dutta.2023@live.rhul.ac.uk} \and Rikke Bjerg Jensen\\ \texttt{ rikke.jensen@royalholloway.ac.uk}}
\affil{Royal Holloway, University of London, UK}
\date{}

\begin{document}

\maketitle

\begin{abstract}
  \noindent We bring to light how some asylum seekers and refugees arriving in the UK experience border control and wider immigration systems, as well as the impact that these have on their subsequent lives in the UK. We do so through participant observation in a support organisation and interviews with caseworkers, asylum seekers and refugees. Specifically, our findings show how the first meeting with the border, combined with a `hostile' immigration system, has a longer-term impact on their sense of belonging. Our observations highlight feelings of insecurity, anxiety and uncertainty that accompanied participants' experiences with immigration systems and processes. We contribute to the growing body of HCI scholarship on the tensions between immigration and (security) technology. In so doing, we point to future directions for participatory and collaborative design practices that centre on the lived experiences and everyday security of asylum seekers and refugees. 
\end{abstract}

\section{Introduction}\label{sec:introduction}

The increasing reliance on technology in UK immigration control has unfolded over the last decade in tandem with policy environments that frame migration as an issue requiring ever tighter forms of control~\cite{miller2016strangers}. The evolution of border control technology rests on an institutionalised ``hostile environment'' framework initiated in the UK in the early 2010s, through daily checks on migrants' access to housing, work and healthcare~\cite{JCWI:hostileenvironment}. This environment also created fertile ground for technological infrastructures that could extend such a verification logic~\cite{YuvalDavis2017}. The requirement for biometric passports, automated e-gates and digital visa processing were entrenched as part of routine UK border governance in the late 2010s and early 2020s~\cite{ozkul2023automating}. This trajectory accelerated after the UK left the European Union (EU) (\emph{Brexit}) in early 2020, with the EU Settlement Scheme\footnote{The EU Settlement Scheme is a scheme for EU citizens and their families to remain in the UK post \emph{Brexit}.} replacing physical documentation with digital-only IDs~\cite{the3million2025digital}. Further, in 2025 e-visas were rolled out to replace physical biometric residence permits. This represents the latest point in a continuum of digital infrastructures being embedded as the default medium for immigration status verification. These developments further suggest how technologies at the border become the operational arm of broader regulatory frameworks that shift the burden of legitimacy and transparency onto migrants as noted in prior work, e.g.~\cite{Dijstelbloem2014, yang2024automateddecisionmakingartificialintelligence}.

Prior work highlights that migrants often experience biometric data collection, digital-only checks and predictive risk assessment not as neutral processes but as encounters that reinforce suspicion and diminish dignity~\cite{KlimburgWitjes2023,Topak2023,CHI:DocBie22}. These effects are particularly acute among asylum seekers and refugees, whose precarious legal positions amplify the stress of navigating opaque and increasingly automated systems~\cite{CHI:DocBie22,SIU:ABGMS23}. The move to e-visas in the UK is especially significant: while framed as an efficiency measure, existing scholarship (e.g.~\cite{madianou2019,Ponzanesi2022,Aizeki2024-wk}) highlights how it risks exacerbating exclusion by rendering immigration status dependent on digital infrastructures that many cannot easily access. 

Our study intervenes at this juncture, documenting how technologically mediated immigration controls are lived and contested~\cite{Leurs2018,Kingston2018}, by capturing the perspectives and  experiences of asylum seekers, refugees and caseworkers navigating UK border control and wider `hostile' immigration systems. This experiential dimension is particularly acute for refugees and asylum seekers as their legal precarity amplifies the stakes of every interaction with opaque systems. 

\paragraph{Contributions.} We conducted a three-month study (June--August 2024) that combined observation at a UK-based charity focusing on immigrant well-being, including practical assistance with immigration-related tasks such as e-visa applications, biometric residence permit queries, GP registration, housing and welfare appointments, while also offering essential items and food distribution. We also conducted interviews with six asylum seekers and refugees (\emph{clients}) and six people working at the charity (\emph{caseworkers, CEO} and \emph{legal aid}). Our study was guided by the following question:

\begin{quote}
	\emph{How do refugees and asylum seekers in the UK experience encounters with border control technologies and the wider immigration system, and how do these encounters shape their longer-term sense of security?}
\end{quote}

\noindent We thus contribute to existing work on how digital infrastructures restructure not just border management but the everyday conditions of life for asylum seekers and refugees~\cite{tazzioli2020making}. We position our work in existing HCI scholarship on borders and technology, while drawing on migration studies and everyday security. Concretely, our work makes four distinct contributions to CHI and, more broadly, HCI scholarship. First, we show how border screening processes were experienced as both hostile and confusing for many asylum seekers and refugees, where they felt treated as a `suspect by default'. We thus extend existing CHI work on border technology in the context of asylum seekers and refugees, which predominantly focuses on pre- and post-border experiences (cf.~\Cref{sec:related-work}). Second, we show how the first meeting with the UK border and associated screening processes shapes feelings of everyday insecurity for many refugees and asylum seekers long after they arrive in the UK. The border itself thus becomes a marker of `otherness' with many participants describing how they felt alienated during their border crossing and subsequent processing; a feeling they said continued even after having been given legal status in the UK. Third, as others have highlighted (e.g.~\cite{coles-kemp2018new,coles-kemp2019accessing,simko2018}), we show the significance of support structures, such as caseworkers, for refugees and asylum seekers navigating digitalised processes of (re)settlement. Fourth, we point to the potential in participatory design interventions and collaborative practices to prepare those fleeing to a new country for what border screening and bureaucratic procedures might entail, while contributing to the re-imagining and re-design of future screening processes that the participants in our study referred to as `de-humanising'. Our work brings to light a tension: many participants imagined the UK as a safe country before arrival, but feelings of anxiety and insecurity underpinned their lives in the UK. 

\paragraph{Foundations.} Our work navigates concepts such as \emph{everyday security}, \emph{belonging} and \emph{hostile environment}, which we define here. We found that for the participants in our study \emph{security} was tied to deeply rooted feelings of uncertainty about their future in the UK and a lack of trust in institutional processes, including border screening and immigration bureaucracy. We thus understand security as relational, through the lived experiences of security; what Crawford and Hutchinson~\cite{BJC:CraHut16} and Nymann~\cite{IPS:Nyman21} refer to as ``everyday security''. This positioning foregrounds how security is shaped by micro and proximate interactions, through temporal (e.g.~security embedded in daily routines and future anxieties), spatial (e.g.~navigating security in and through different physical spaces) and affective (e.g.~emotional responses to security decisions and practices), rather than macro and institutional frameworks. These ``quotidian aspects of social life''~\cite[p.~2]{BJC:CraHut16} further shaped participants' feelings of not \emph{belonging} in the UK, which were rooted in experiences of being labelled ``suspect by default'' through both border screening and wider immigration processes. We understand belonging as a multidimensional construct encompassing affective attachment to place, social recognition by others and a sense of legitimate presence within a community or country~\cite{Yuval-Davis2011-mk}. Our work thus points to the fragility of belonging within the UK policy context of what has been termed a ``hostile environment'' (later rebranded as the ``compliant environment''). Introduced through the UK Immigration Act 2014 and expanded in 2016, these policies extended immigration enforcement into everyday life by requiring landlords, employers, healthcare staff, banks and other service providers to verify immigration status~\cite{YuvalDavis2017}. This framework created what scholars describe as a ``culture of disbelief'' in which migrants bear continuous burden of proving legitimacy through documentation~\cite{Khosravi2010-kd}.

\section{Related Work}\label{sec:related-work} 

We synthesise two converging strands of scholarship on \emph{digital border regimes} and the impact on migrant experiences. In~\Cref{sec:hci-border-tech}, we first position our work within HCI scholarship, before outlining existing work on immigrant experiences of borders in~\Cref{sec:immigrant-experiences-borders}. Collectively, these literatures reveal that while digital border infrastructures promise efficiency and security, they also reproduce hierarchies of mobility and knowledge.

\subsection{Borders and Technology in HCI}\label{sec:hci-border-tech}

HCI research has increasingly turned its attention to how border control technologies mediate migration~\cite{talhouk2016,sabie2019}. Rather than treating borders as fixed geographic lines, some HCI scholars frame them as sociotechnical infrastructures that gather data, make classifications and decide who can cross and under what conditions~\cite{razaq2025,suchman2007}. Central to this understanding is Chouliaraki and Georgiou's~\cite{chouliaraki2019} `Digital Border Assemblages (DBA) framework', which shows how borders combine both technical systems (such as biometrics and databases) and symbolic narratives (such as national identity and security discourses) to regulate migrant mobility.
It thus (re)frames borders as everyday systems of control that extend far beyond the checkpoint itself as also highlighted by Madianou~\cite{madianou2019}. Building on this framework, a growing body of HCI scholarship demonstrates how technologies such as biometric scanners or digital visa platforms not only process data, but also create social hierarchies -- privileging some migrants while excluding others~\cite{razaq2025}. For instance, in research with newcomers in Sweden, the authors of~\cite{coles-kemp2018new},~\cite{coles-kemp2019accessing} and~\cite{jensen2020civic} showed how digitised processes directly shaped refugees' resettlement experiences that required interaction with opaque, automated systems.

Alongside these structural concerns, privacy and surveillance have emerged as critical dimensions in broader computer security and technology scholarship focusing on refugees and migration. The authors of~\cite{guberek2018}, for instance, showed that undocumented migrants adopted `security-through-opacity' practices, including selective disclosure, pseudonymous accounts and channel-switching to reduce risks during routine communication. In their study on the security needs of refugees in the US, the authors of~\cite{simko2018} found that a greater reliance on technology not only heightened the significance of computer security, but also that the reliance on digitised services was often experienced as barriers to a successful resettlement because they required technology familiarity \emph{as well as} language and cultural familiarity. Similarly, in~\cite{steinbrink2021} Steinbrink investigated asylum seekers' privacy perceptions during migration, showing that those who fled government persecution often developed sophisticated mental models of surveillance risks that could affect their asylum status. For example, they employed various strategies including maintaining anonymity, modifying communication patterns and sometimes avoiding the use of mobile phones~\cite{steinbrink2021}. These studies underline that digital border infrastructures do not make institutions more transparent; instead, they shift the burden of legibility onto migrants, who must continuously disclose data, adapt to automated systems and perform compliance to be recognised as legitimate travellers or residents. As Dijstelbloem and Broeders~\cite{dijstelbloem2015border} demonstrate, biometric technologies operationalise this burden-shifting: while states collect ever more granular data about migrants' bodies and movements, the decision-making processes behind visa approvals, risk assessments and status determinations remain opaque and inaccessible to those being assessed. This asymmetry -- that is, \emph{compulsory transparency} for migrants coupled with \emph{institutional opacity} -- creates conditions where, as our findings show (cf.~\Cref{sec:findings}), refugees and asylum seekers describe the screening experience as simultaneously invasive and incomprehensible.

\subsection{Experiences of Border Processing}\label{sec:immigrant-experiences-borders}

A broad range of migration studies have examined border (control) technologies from the perspective of migrants themselves. These studies highlight the agency, resistance and embodied nature of border encounters. In this section, we bring such literature into conversation with HCI scholarship. 

On the one hand, prior work shows how migrants actively resist technological control. Metcalfe's~\cite{metcalfe2022} ethnographic study in Greece documents how migrants developed strategies to resist biometric systems, such as altering their fingerprints, to contest the rigid allocation of asylum claims under the `Dublin System'.\footnote{This is a system that assigns claims to the first country of entry.} This resistance connects to Scheel's~\cite{scheel2013embodied,scheel2013autonomy} autonomy of migration perspective, which shows how even under tight surveillance, migrants often find ways to sustain mobility and agency. On the other hand, these same technologies create new forms of conditional citizenship. Lemberg-Pedersen and Haioty's~\cite{lembergpedersen2020} work with Syrian refugees in Jordan shows how biometric registration produced ``quasi-citizenship'', where access to services depended on surrendering biometric data. In a similar vein, Topak's~\cite{topak2019border} study with migrants in Greece identified multiple forms of migrant subjectivity, including religious, nomadic, abject and dissident. Apparent in this work is how migrants not only comply or resist border technologies, but also negotiate complex identities shaped by increasingly technologically mediated border systems.

Central to these experiences is how they become \emph{embodied encounters}. Migrants interact with technologies such as fingerprint and iris recognition scanners through their bodies, and these interactions shape their mobility and identity~\cite{scheel2013embodied}. Thus, these encounters are not just technological but emotional -- they are \emph{felt}. For example, Borrelli~\cite{borrelli2021} shows how negotiations between migrants and \emph{bureaucrats} are mediated by affect, frustration, fear and sometimes empathy, making border control a deeply emotional process for many. Further, as noted by Bannerman~\cite{bannerman2019}, technologies designed for the purpose of monitoring are not limited to the experience of the individual but affects their social networks. This relational view brings to light how border surveillance practices also transform families, communities and diaspora connections into sites of monitoring and control~\cite{razaq2025}. Thus, encountering an opaque biometric system at the border may generate anxieties, shape future digital practices and condition how migrants engage with institutions after arrival~\cite{fisher2020}. Therefore, the operation of border systems rely on externalising the work of transparency: migrants -- not institutions -- must repeatedly prove identity, intent and credibility, often through bodily compliance, repeated data submission and ongoing emotional labour during encounters with bureaucrats. 

Lemberg-Pedersen and Haioty~\cite{lembergpedersen2020} term this dynamic ``surveillable refugee bodies'', where biometric registration transforms refugees into legible subjects whose access to services depends on continuous bodily disclosure. Similarly, Metcalfe's~\cite{metcalfe2022} documentation of fingerprint resistance demonstrates migrants' awareness that once biometric data enter systems, they lose control over how that information circulates or shapes their futures. These bodies of work show that as border technologies become more automated and data-driven, institutional accountability recedes, while migrants shoulder increasing responsibility to make themselves interpretable to systems that often remain inaccessible.

\paragraph{Research Gap.} Bringing HCI and migration scholarship into conversation highlights how border control technologies are not passive tools of administration; they actively produce subjectivities, hierarchies and emotions that shape the migrant journey~\cite{madianou2019}. Research has documented the `chilling effects' of government surveillance, where awareness of monitoring deters legal activities and reduces political participation among targeted communities~\cite{penney2016,stoycheff2019}. This suggests that the impact of border control technologies and immigration systems extends far beyond the moment of crossing, shaping migrants' civic engagement and sense of belonging long after arrival in their host country. While HCI research has often concentrated on post-arrival (digital) experiences, migration studies tend to focus on the moment of border crossing. Thus, how these initial border encounters shape and structure migrants' everyday security~\cite{IPS:Nyman21,BJC:CraHut16} and belonging in their host countries remains under-explored. We take one step in bridging this gap. We focus on how refugees and asylum seekers experience UK borders and how they reflect on their experiences after arriving in the UK.

\section{Methodology}\label{sec:methodology}

Our methodology combines participant observation at a UK-based charity and semi-structured interviews with individuals who visited this charity for support, i.e.~`clients', including asylum seekers and refugees, as well as caseworkers who worked at the charity. We established contact with this charity through networks connected to refugee and immigrant welfare in London, UK. We negotiated initial access through the CEO who invited us to visit their centre, where the researcher was also introduced to several caseworkers. All caseworkers were also provided with an information sheet during this visit, before accepting the researcher into their workspace. 

Everyday engagements during the three-month study was shaped by interactions with caseworkers who acted both as gatekeepers to clients and, in some cases, as interview participants (cf.~\Cref{tab:participants}). In particular, caseworkers introduced the researcher to clients who they felt would be comfortable interacting with the researcher. Thus, we relied on the caseworkers' expertise and experience to facilitate initial contact. The researcher stressed the voluntary nature of participation in all engagements, while highlighting that non-participation would have no impact on their access to charity support. This was particularly important as clients were structurally dependent on the charity for legal guidance, essential services and emotional support. Here, it is also worth noting that the specific charity had previously hosted academic visitors and the presence of a researcher was not unfamiliar to those at the charity.   

\subsection{Participant Observation}\label{sec:participant-observation}

Participant observation enabled us to engage with asylum seekers and refugees in an environment where they felt safe and supported. Specifically, one researcher (cf.~\Cref{sec:positionality}) spent three months -- June, July and August 2024 -- with one charity focusing on supporting immigrants arriving and settling in the UK. This included practical support and advice such as assistance with e-visa applications, biometric residence permit queries, doctor's registration, housing concerns and welfare-related appointments. Throughout the three months, the researcher observed day-to-day activities at the charity, engaged with people who came to the charity looking for advice, to receive essential items and food, and to attend appointments as well as interactions between charity staff and clients as they navigated a wide range of immigration-related concerns and questions. These were often handled under significant time pressure, requiring staff to balance procedural expertise with the ability to offer reassurance and emotional support. The charity space was separated on two floors, with office work taking place on the upper floor and the first floor being used for distributing food rations and other essential items. The researcher was primarily located on the upper floor to avoid disrupting aid distribution. Observational work was thus restricted to spaces where everyone was informed and consented to the researchers' presence. Full field notes were taken throughout the fieldwork~\cite{EmeFreSha11}, capturing daily observations and the fieldworker's initial reflections and interpretations.

During the first month, the researcher aimed to integrate herself in the daily activities, becoming more and more familiar with the processes and daily interactions and ensuring that people at the charity were aware of and agreed with her presence (cf.~\Cref{sec:ethics}). Immersion in the day-to-day work of the charity enabled insights into the daily hurdles of immigration work, the frustrations inherent in navigating inflexible bureaucratic processes and the helplessness many felt in this context. These frustrations were exemplified by the CEO, who explained that people fleeing the war in Ukraine were offered refugee status through special Government schemes that provided immediate assistance such as free housing and food -- what they described as a ``red-carpet welcome''. In contrast, people from Afghanistan and several African countries often faced protracted legal struggles to be recognised as refugees, leaving them with limited or no access to comparable State support. Similar examples were shared by others which led to a feeling that immigrants from some countries were considered ``more deserving'' within the UK immigration system. 

\subsection{Semi-Structured Interviews}\label{sec:interviews-method} 

Semi-structured interviews were chosen due to their exploratory nature; they are sufficiently structured to provide consistency across interviews and to address particular research questions, while leaving space for participants to offer new meaning to the topics (e.g.~\cite{NYU:GalCro13}). All interviews were conducted \emph{in situ} by the researcher in July and August 2024, and involved immigrants who periodically visited the charity for different reasons, as well as caseworkers (cf.~\Cref{tab:participants}). 

At the start of each interview, the researcher explained the study purpose and interview process, clarified that participants were not obliged to take part or that the support they received from the charity was not contingent on them participating in the study. Power asymmetries were further shaped by language mediation, where caseworkers translated when required to ensure that all aspects of the study and what participation involved were understood by participants. Caseworkers were further present during interviews to both translate where needed and to reassure clients, yet they were not actively participating in the interview or steered conversations. The presence of a caseworker during interviews was agreed at the start of each interview and the researcher made sure that the participant felt comfortable with this arrangement; it was noticeable that many participants felt more comfortable engaging in the research with the caseworker present. Interviews with caseworkers also took place during the researcher's time at the charity. No interview was audio recorded on the advice of the charity. Instead, the researcher took hand-written notes, which were expanded into detailed descriptions the same day, following the guidelines provided in~\cite{QR:RMBKTMS20}. Interviews loosely followed an interview guide (cf.~\Cref{sec:interview-guides}).

The role of caseworkers as both translators, mediators and gatekeepers inevitably shaped the production of data, even though we tried to ensure that the dominant voices in each interview were those being interviewed. As \emph{gatekeepers}, caseworkers introduced the researcher to clients based on perceived comfort. The sample is thus partly constructed through their judgements of who was `suitable' to participate. As \emph{translators}, caseworkers enabled communication but also selectively filtered meaning; for example, pauses, emotional hesitations or culturally embedded expressions were sometimes condensed into brief English summaries. Hence, interview notes were sometimes not verbatim accounts. As \emph{mediators}, during interviews to support and reassure participants, they may also subtly have redirected the emotional tone of some interviews. While there is no indication in our data of caseworkers intentionally altering what was expressed by other participants, their presence in interviews inevitably created a narrative layer in which clients' experiences were interwoven with caseworkers' interpretive practices, professional responsibilities and protective assurances. Further, the relationship between the caseworkers and their clients is one of uneven power, which could have shaped how asylum seekers and refugees chose to respond to our questions and which stories to share. As \emph{interviewees}, caseworkers used first person narration when drawing on their own personal experience and third person when drawing on that of clients. Both their personal and professional insights significantly enriched our data and understanding. Thus, the voices captured in the study are best understood as co-produced accounts, dominated by refugee and asylum-seeker experiences but shaped by caseworkers translations and mediations and the researchers' positional reading of both (cf.~\Cref{sec:positionality}). 

\begin{table*}[t]
	\centering 
		\caption{Overview of interview participants.}
		\resizebox{\textwidth}{!}{\begin{tabular}{cccccc}
			\toprule
			(A) ID & (B) Home Country & (C) Designation & (D) Current Status in UK & (E) Time in UK &  Interview Duration\\
			\midrule
			AS1 & Palestine & Client & Asylum Seeker  &5-9 years & 55 min\\ 
			AS2 & Iran & Client& Asylum Seeker &2-4years & 65 min\\
			R1 & Bangladesh & Client &Refugee &2-4 years & 61 min\\
			R2 & Afghanistan  & Client & Refugee & 0-1year & 71 min\\
			R3 & Nigeria & Client & Refugee & 5-9years& 67 min\\
			M1 & India & Client & Migrant &2-4years & 65 min\\
			& & & (Awaiting Visa Status) & & \\
			CW1 & Afghanistan &  CEO & British Citizen & 30+ years & 50 min\\
			CW2 & Afghanistan &  General Assistant & British Citizen & 10+ years & 40 min\\
			CW3 & Syria & Caseworker & British Citizen & 5-9years & 30 min\\
			CW4 & Afghanistan & Caseworker & British Citizen & 5-9years & 65 min\\
            CW5 & Kenya & Legal Aid & British Citizen & 5-9years & 40 min\\
            & & (OISC Level One and Two) & & & \\
			CW6 & UK & Caseworker/Administrator & British Citizen & Born in the UK & 30 min\\
			\bottomrule
		\end{tabular}}\par
	\label{tab:participants}
\end{table*}

\subsection{Data Analysis}\label{sec:analysis}
We analysed the complete dataset through multiple rounds of coding and engaged in collaborative analysis. 

\subsubsection{Coding Rounds}\label{sec:coding-rounds}

We conducted a reflexive thematic analysis following Braun and Clarke~\cite{QRP:BraCla06,QRSEH:BraCla19} to understand how refugees, asylum seekers and caseworkers engaged with UK border control and immigration systems. This interpretive stance allowed us to trace both shared patterns and individual experiences across the interviews. The primary researcher first cleaned up all interview and field notes before familiarising herself with them through a full read-through. Our analysis then unfolded in three coding rounds. First, the primary researcher conducted open coding on all interview and field notes, generating initial descriptive labels to reflect this coding. This captured emotional, institutional or technological moments such as ``\textit{retina scan at Heathrow}'', ``\textit{application got stuck}'', ``\textit{fear of being rejected}'' or ``\textit{borrowed a phone to apply}''. In the second round, both authors reviewed and discussed the initial codes, clustering them into higher-order concepts. The third round involved refining these concepts into a stable set of cross-cutting themes reflected in~\Cref{sec:findings}.

\subsubsection{Disagreement Resolution}\label{sec:analysis-disagreements} 

Observational data often provided contextual grounding for interview data (e.g.~caseworkers' reassurance practices, client anxiety around paperwork), while interviews clarified meanings behind behaviours observed at the charity. Reading these two forms of data `against' each other in collaborative sessions helped to identify convergences and tensions. Points of disagreement were resolved by first returning to the original data to ensure that final category assignments remained grounded in participants' words. Generally, agreement was reached during the same session. In the few instances where this was not the case, codes were temporarily retained as multiple possibilities and revisited in the next session with fresh readings until agreement was achieved. We then constructed summaries that captured our resolve and the concepts. We include an example of our reflexive coding table in~\Cref{sec:reflexive-coding-table}.

Given our different positionalities (cf.~\Cref{sec:positionality}), we took great care in reading across the constructed themes, challenging our interpretations and initial coding. Through these analytical discussions we also observed how asylum seekers and refugees often described border technologies as ``\textit{scary and confusing}'', combining unfamiliar interfaces with their legal precarity. Caseworkers, many with immigration histories themselves, reported emotional exhaustion from helping others navigate processes they once struggled with. 

\subsubsection{Positionality}\label{sec:positionality}

Our individual and collective identities shaped how we conducted the research and interpreted the data. Rather than `neutral observers', we adopted a position of `situated researchers'. The researcher (first author) is a South Asian migrant with prior exposure to UK border systems, which shaped both participant rapport and data interpretation. The second author is a European and an immigrant with settled status in the UK, researching the security needs of groups in higher-risk contexts. We recognise that our differentiated experiences of immigration systems shaped both our research engagements and our interpretations; yet, we strove to remain situated in the context of the study and employed a reflexive analytical approach to continuously move between the data and analytical discussions. 

\subsection{Ethical Considerations}\label{sec:ethics}

Our study was approved for self-certification through our institutional Research Ethics Committee (REC). We designed our study in a way that minimised the collection of personally identifiable information at all stages of the research, and no interview or field notes included this information. We adapted our approach to obtaining informed consent to suit the participants and the context of the research. Because of language barriers between refugee and asylum-seeker participants and the researcher, caseworkers were relied upon to translate the information sheet and consent form to ensure that its content was fully understood. Thus, the researcher provided caseworkers with information sheets at the start of the research both to ensure they were aware of the researcher's presence and that any participation was voluntary. Many participants held deep-seated apprehensions towards signing formal documents. For example, one participant signed their consent form in their native language, while another took additional steps to verify the contents before signing. Some used only initials or altered names to further ensure their anonymity. Thus, informed consent procedures highlighted deep-rooted feelings of vulnerability and insecurity among participants, in light of their experiences of bureaucratic processes and official paperwork, illiteracy or past trauma. To safeguard their privacy, all (given) names were anonymised in the interview scripts and replaced with alphanumeric codes. We do not name the charity we worked with or their specific location to protect participants who took part in our study. For similar reasons, we do not report participant gender, but note that both women and men took part in the research, while we cannot make our dataset publicly available as it would not be possible to fully anonymise participants and their contributions. 

\section{Findings}\label{sec:findings}

We outline our findings in line with our analysis (cf.~\Cref{sec:analysis}). While distinguishing between border screening and broader immigration processes, we show how they operate as a continuous sociotechnical system. We define border screening technologies as technological components of the initial border encounter, such as biometric fingerprinting, iris scans, automated identity checks, digital visa verification and the backend systems that process this data. These systems form the first node in a larger chain of immigration controls but do not, on their own, explain the protracted insecurity participants described. The broader immigration regime encompasses the full procedural apparatus that follows the initial encounter, including immigration officer questioning, case assessments, status verification, documentation updates and e-visa transitions, to mention a few. Drawing on critical border studies~\cite{broeders2007,dijstelbloem2015border}, we thus recognise  border technologies as extending control far beyond the checkpoint itself. For instance, Broeders~\cite{broeders2007} highlights that in the European context, digital borders operate through networked databases that enable ongoing surveillance and ``internal control'' of migrants long after initial entry. 

Our analytical transition from the initial border checkpoint to these downstream processes is therefore intentional: participants described experiencing these elements as a single, extended infrastructure of control, where the confusion and fear triggered by early technological encounters shaped their expectations, trust and emotional responses throughout the much longer bureaucratic journey. First,~\Cref{sec:meeting-the-border} exemplifies how participants experienced being processed at UK borders, gleaned from interviews. Second, in~\Cref{sec:across-border}, we report on the longer-term effects on the sense of belonging described by participants as they tried to make a life in the UK. 

\subsection{Meeting the Border}\label{sec:meeting-the-border}

Many participants described how crossing into the UK was fraught with emotional turmoil that influenced their first `meeting' with the country. Their perspectives were often grounded in limited prior exposure to screening technologies and the urgency of their migration. Although they viewed the UK as a `safe country', their encounters with border control technologies contradicted their views of what constituted safety and security. Distinct from the wider immigration system, these point-of-entry technologies had made several participants uneasy about the use of personal data. 

\subsubsection{Fear in the Face of Faceless Systems}\label{sec:fear-tech-border}

Many participants shared how arriving in the UK and encountering different forms of border control systems had led to heightened uncertainty. This uncertainty was tied not only to unfamiliar technologies such as fingerprinting, facial photography, body searches and iris scans -- but also to personal histories that shaped how these systems were interpreted. Participants who had fled authoritarian regimes or conflict emphasised that giving up personal information was tied to `danger', making the demand for data at the border particularly distressing. Many described feeling they had surrendered ``too much personal freedom for the promise of safety'', moving from escaping threats in their home countries to facing intense monitoring in the UK.

R2, who fled Afghanistan by boat with two young children, recalled being fingerprinted, photographed and scanned immediately upon arrival. Exhausted and unable to speak English, R2 felt ``disorientated'', anxious about answering questions correctly and unsure why the technologies were being used: ``What will they ask? What if I don't know the correct answer?''. For many like R2, the urgent need to secure protection overshadowed concerns about what information was being captured. R3, an engineer from Nigeria, shared how they had arrived believing a legitimate job was waiting for them but had learned at a UK airport that the employer did not exist. Having spent their savings on the journey, R3 felt ``intimidated'' during fingerprinting, iris scanning and document checks: ``I had done everything right, or so I thought''. R3 described the fear that the biometric data collected could later be used against them, especially as they were stranded and without support.

Like R3, M1 arrived in the UK from India with what they thought was a job opportunity but learned, when arriving at the UK border, that this was a scam. M1 was left in a legal limbo and when their documents were scrutinised and their biometrics taken, they felt scared about what would happen to them. M1 ``had never experienced anything like this before'' and shared how they were startled to have their eyes scanned and unsure what information was being captured. M1 recalled: ``What if something goes wrong with the machine?''. Many participants had fled authoritarian regimes and shared how they had tried to convince themselves that providing the required information and being processed through \emph{these} ``machines'' was safe: ``this is the UK, it must be safe'' (M1). R3, however, highlighted how each `click' of the camera and scan of their fingers amplified a feeling of dread; that ``a faceless system'' would determine their future in this country. They worried that an error or misunderstanding could result in them being detained or sent back. Recognising that the process was out of their control, R3 noted: ``The process taught me that you could prepare for everything but something might still go wrong.'' 

Nearly all participants recalled the border screening as an intense and nerve-wracking experience. Refugees and asylum seekers often arrived already carrying trauma from their home countries or their journeys, and the border checks often compounded their anxiety. R2 described being overwhelmed by confusion and fear when they were unable to speak the language and faced a series of questions from border officials.

\begin{quote}
	``Everything was so different and confusing [\ldots] It made me feel like I had to prove that we were genuinely seeking safety and not a threat. It was stressful, especially with my children watching.'' (R2)
\end{quote}

\noindent Like others, R2 had struggled to understand what was being asked of them and feared not being believed when explaining their need for refuge. This was accompanied by an uncertainty of not knowing whether they would be allowed to stay. R3 shared how the screening process ``was efficient, but it made me feel like I was entering a very controlled world''. While many participants arrived in the UK in the belief that it was a safe place, their experiences at the border -- of being screened and processed -- often conflicted this belief. 

\subsubsection{``The whole procedure felt dehumanising''}\label{sec:dehumanising}

A common thread across our data was the feeling of being treated with suspicion rather than compassion, leaving many refugees and asylum seekers distressed from the moment they encountered the border. Participants described how the process felt ``impersonal and interrogative''. Like R3, many felt ``just another case, not a person''. AS1, who arrived from Afghanistan by boat, recalled that their first contact with UK authorities occurred inland rather than at the port of entry. They described the fingerprint and face scanning as ``unsettling''. For AS1, ``the whole procedure felt dehumanising'', with their personal story feeling secondary to whether their data ``checked out''. Similarly, R3 explained that the automated systems at the border ``contributed to a sense of depersonalisation'' and noting that ``there's no one to ask, no one to reassure you in the moment'' (FN).\footnote{\emph{FN} denotes `field notes'}

Unlike R3 and AS1, R1's, a Bangladeshi refugee, experience when meeting the border had been one of patience, where R1 felt that they were being listened to and were given the opportunity to clarify any answers they had given: 

\begin{quote}
	``They asked me a lot of questions [\ldots] It was thorough, but I think it was fair. I'm happy now because I got my refugee status in about a year, which was a huge relief.'' (R1)
\end{quote}
 
\noindent R1 recalled how the official who handled their case and interviewed them showed a ``degree of empathy'' which made ``the difficult questions easier to bear''. R1 further stressed how lucky they had been as ``the majority of people have had distressing interactions with officials''. Many refugees and asylum seekers encountered during the fieldwork described being treated as ``potential criminals'' (FN) and that they were ``guilty until proven innocent'' (FN). Some caseworkers underscored this view, with CW5, a caseworker originally from Ghana, criticising what they described as ``a presumption of guilt underlying the whole asylum system''. CW5 noted that while data collection and vetting were necessary, the screening process often treated refugees and asylum seekers untrustworthy by default:

\begin{quote}
	``The [UK] Home Office\footnote{The Home Office is the UK Government department responsible for crime, the police, drugs policy and immigration and counter terrorism.} operates under the assumption that everyone might be lying, and it's up to the asylum seekers to prove otherwise. This whole process can strip away their dignity.'' (CW5)
\end{quote}
   
\noindent CW5 exemplified what they considered to be more stringent document checks, interviews and data logging, which created an ``atmosphere of suspicion'' from the moment refugees and asylum seekers first interacted with the UK border. The assurances about data privacy (such as signing consent forms or being told about data protection laws) were described as doing little to comfort people who were often in a state of crisis and not fluent in English when they arrived at the UK border: ``These rules are not for the safety of the migrants; they're for consolidating the position of the organisation as legitimate'' (CW5). 

\subsection{Beyond the Border}\label{sec:across-border}

Our data shows how the anxiety and uncertainty that shaped participants' first encounters with the UK border and the technologies used to mediate it, continued to shape their experiences long after their first encounter. In this section, we present findings related to caseworker perspective and experiences.

\subsubsection{``Smaller cogs in the system''}\label{sec:caseworkers}

Through interactions with caseworkers during the study, we observed a tension in how they conceived of border control systems. On the one hand, they considered robust border screening processes necessary for ``public safety and fraud prevention'' (FN) while, on the other, witnessed how those same systems instilled fear and confusion in the people they supported every day. CW2, who assisted with legal paperwork at the charity, gave concrete examples of why they believed that certain border control processes were beneficial:

\begin{quote}
	``I know of cases where smugglers or even individuals involved in war crimes were caught because of these technologies [\ldots] These systems make the UK safer for everyone, including those of us who have come here seeking refuge.''(CW2)
\end{quote}

\noindent Similar views were expressed by other caseworkers who noted that the safety both of those arriving in the UK as immigrants and of the systems themselves relied on robust border screening processes as well as biometric databases and data sharing between countries. This view, however, was contrasted by several refugees and asylum seekers as exemplified in~\Cref{sec:meeting-the-border}; the very same systems that were seen to inculcate trust from the perspectives of some caseworkers (e.g.~CW2), were experienced as ``dehumanising'' by most refugees and asylum seekers. Thus, our data points to a tension between personal privacy and collective security. 

CW3, a Syrian refugee-turned-support-worker, stressed that increasing automation at UK borders could increase stress among refugees and asylum seekers who were often in a ``fragile state of mind'' when they arrived in the UK:

\begin{quote}
	``[A] metal detector or a biometric kiosk cannot comfort a terrified family that has just escaped death.Technology is crucial for security and fairness, yes, but it can't understand a trembling mother or a confused child.'' (CW3) 
\end{quote} 

\noindent Underscoring the feelings expressed by several refugees and asylum seekers, CW4, originally from Afghanistan, noted that while many did not understand how their personal information would be used they felt unable to refuse biometric checks or interviews. CW4 explained that many asylum seekers often complied without asking questions because they feared this would  jeopardise their claim. However, our data showed that most participants we engaged trough the study had little to no knowledge of why certain details were being collected or what would happen to such details.

CW1 felt a ``constant pressure to maintain composure and provide reassurance'' to newly arrived clients going through the asylum process, while CW3 voiced frustration at the system's inflexibility: ``the focus on efficiency and strict protocols leaves little room for addressing individual needs''. They gave the example of a client who, due to a past trauma, had become distressed by needing to have their fingerprints taken, yet, they had no choice but to encourage them to comply with the system. CW3 highlighted: ``Sometimes I feel like my hands are tied by these procedures.'' CW2 recounted an example of how they had explained the purpose of a fingerprint scanner to a scared asylum seeker who encountered this technology for the first time. CW2 explained that because they could only support refugees and asylum seekers after they had crossed the UK border,\footnote{Everyone we spoke to at the charity underscored how they were only allowed to offer support once an individual had arrived in the UK.} they were ``smaller cogs in the system''.

\begin{quote}
	``By the time we meet them, many are already shaken. We do our best to explain and comfort, but we can't change that first impression they got.'' (CW2)
\end{quote}

\noindent Through our analysis it also became clear that caseworkers and the organisational setting of the charity played a decisive mediating role in shaping how refugees and asylum seekers understood, processed and coped with the UK immigration system. Caseworkers often functioned as the extension of border interfaces, not a peripheral `add-on'. In particular, refugee and asylum seeker participants consistently reported that clarity and reassurance only materialised \emph{after} the checkpoint -- at the charity desk, where caseworkers translated letters, explained biometric procedures and rehearsed appointments. Thus, the `interaction surface' of border technologies extends beyond kiosks and counters into support organisations. This extension is not rhetorical: comprehension of what a fingerprint or iris capture means is produced through caseworker explanations, printed examples and step-by-step walk-throughs. Thurs, caseworkers, four of whom (CW1, CW2, CW3, CW4) were former refugees, were often the first to explain the purpose of these measures. CW2, for example, routinely reassured clients that ``the scan is just the system anchoring [your] identity'' and helped to contextualise technological processes within the broader legal framework. Our data similarly shows how clients' interpretations of border technologies were often inseparable from caseworker explanations, which filled in gaps left by opaque or confusing state systems. 

Building on this foundation, our study uncovers three interlocking types of work performed by caseworkers. First, \textit{explanatory work} involves constructing mental models for clients -- ``what the scanner records'', ``where data goes'', ``why refusal is treated as non-cooperation'' -- often by anchoring these in familiar analogies. Second, \textit{affective work} involves actively regulating emotion -- slowing the pace, validating fear and scripting self-talk for interviews. Third, \textit{procedural work} includes pre-filling forms, collecting documents, scheduling biometric appointments and simulating kiosk flows. These steps convert rule-texts into ordered actions and decrease error cascades -- ``application got stuck'', ``wrong document type'' -- a pattern repeatedly captured in our field notes.

At the same time, our findings show that while caseworker discretion was \emph{real} it was also \emph{bounded}. Caseworkers could not waive biometrics or alter outcomes, but their discretion was captured in how they could frame, pace and rehearse requirements. This bounded discretion highlights a tension in our data: caseworkers humanise encounters (clients were seen to cope better), yet they experience ``hands-tied'' frustration and exhaustion because structural conditions remain unchanged. This boundedness also meant that caseworker labour risked becoming a compliance infrastructure; by smoothing frictions and repairing trust they enabled uptake of the very systems clients found alienating. 

\subsubsection{Memories of the Border}\label{sec:memories-border}

The initial feelings of alienation and dehumanisation were not left at the border, with several participants sharing how it had been hard to shake off the memory of that first encounter. The anxiety that many participants described having felt at the border also shaped their continued uncertainty about their legal status and the bureaucratic processes that followed. Several asylum seekers explained how, after the initial screening, they entered a protracted period of waiting and continuous interviews, check-ins and evidence submissions. Many shared how this process had strained their mental health. For example, AS1 spent three years in a legal limbo, awaiting a decision on whether they would be granted asylum in the UK. During that time, the initial fear they experienced at the border was transformed into a chronic state of uncertainty. While living in poor housing and unable to legally work, AS1 shared how they were constantly worried about their future in the UK and their family's fate in Palestine.

AS2 similarly described their daily life as a mix of hope and worry. After formally seeking asylum, AS2 explained how they woke up each day not knowing if or when they might be detained or forced to leave. For AS2, the extensive data they provided during the asylum application process, including fingerprints and personal details meant to support their case, had left AS2 feeling increasingly vulnerable and exposed to a system that they neither knew nor understood. AS2 joked that their ``fingerprints are probably travelling more places than I am'', referring to how their data was shared and checked across agencies while AS2 remained stuck in a legal limbo. 

Participants with comparatively stable statuses also shared how the rigorous entry process had left a lasting imprint of insecurity. R1, a refugee from Bangladesh who had recently obtained refugee status, explained how their relief at reaching what they considered to be a safe country meant they had not questioned what data was being collected: 

\begin{quote}
	``Honestly, I don't know much about personal data or why it's important. I didn't ask many questions \ldots I was mostly focused on getting my refugee status. As long as I can live and work here, I'm happy.'' (R1)
\end{quote}

\noindent Several asylum seekers emphasised that they had to sacrifice privacy given their precarious circumstances. AS1, who lived without family or legal status, described the years following their arrival as marked by ``hopelessness and fear'', explaining that the priority had simply been to remain in the UK and they had therefore submitted to any process required for asylum. AS2 recalled how, after two years in hiding and working in informal jobs, they were ``hopeful but also anxious'' about obtaining asylum, noting: ``I will tell them anything they want to hear if it means getting refugee status.'' They described their patience wearing thin after years of compliance with every requirement.

Some long-term refugees and naturalised British citizens (now working as caseworkers) shared how they still felt anxiety during any encounter with border control systems. CW1, who had lived in the UK for 30 years after fleeing conflict, shared: ``Even now, when I go through some of these screenings, I can't help but feel a sense of insecurity and like an outsider.'' Further, CW2, who arrived in the UK as a refugee but had become a naturalised British citizen, shared that while they trusted and understood the need for security systems, ``there are moments \ldots where I feel like I'm back to being that scared newcomer, even though I've been here decades''. Thus, many shared how the memories of the precarity they felt when they initially arrived in the UK became a feature of their daily lives and everyday (in)security.

\section{Discussion}\label{sec:discussion}

We consider our study a point of departure for exploring the (longer-term) impact of border control and immigration systems on those whose lives often depend on being processed through them. Here, we discuss our findings (cf.\Cref{sec:findings}) in light of prior work across HCI and migration studies, as well as everyday security, to both position our work in these wider discourses and to contribute empirical insights to their theoretical foundations. In \Cref{sec:racialised-tech} we examine how border technologies function as sites of othering, showing how screening processes produce lasting feelings of insecurity and surveillance. In~\Cref{sec:implications} we consider the implications of our findings for HCI research, proposing design interventions that centre on agency restoration both before and at the border. Lastly, we explore the challenges inherent in participatory design within border control contexts in~\Cref{sec:design-challenges}.

\subsection{Identity and Othering through the Border}\label{sec:racialised-tech}

Our findings suggest how border crossing and the experiences of and with screening processes in this context are deeply intertwined with individual and collective identities. Several participants in our study expressed how they experienced being treated as a ``suspect by default'' and how the screening processes exacerbated this feeling; often referred to as ``dehumanising'' and ``depersonalised'' (cf.~\Cref{sec:dehumanising}). Participants experienced both language and cultural barriers in their interactions with the technologies and immigration systems that they considered to be deciding their future in the UK. We are not the first to highlight relationship between identity, security and border control infrastructures. Existing work brings to light the concept of \emph{racialised surveillance}, with Browne~\cite{browne2015} defining racialised surveillance as ``enactments that reify boundaries along racial lines''. This suggests how border control systems operate as mechanisms of social control, defining what is ``in or out of place'' based on racial categories~\cite{razaq2025}. The participants in our study had experienced the border screening processes as both confusing and scary, while they had to prove that their need for refuge was ``genuine'' (cf.~\Cref{sec:fear-tech-border}). To them, the UK border and the technologies that sorted who was `in' and who was `out' of place were not benign tools of administration. Instead, they presented as active infrastructures that determined whether they would be allowed to enter the country or not; thus, embedding structural hierarchies of grouping and sorting~\cite{madianou2019} that extended into their future existence in the UK. 

Our findings show that such dynamics of `othering' and `racialised surveillance' are not limited to biometric checkpoints but reverberate across an entire immigration system that refugees and asylum seekers must (successfully) navigate over time. Participants in our study described this not only at the moment of screening but as an enduring condition that shaped later interactions with welfare offices, job centres and subsequent border crossings (cf.~\Cref{sec:across-border}). Our work thus shines a light on the temporal and spatial conditions that produced differing forms of `othering'; extending existing HCI work into digital borders, e.g.~\cite{razaq2025}.

Instead of security, several participants shared how the border met them with procedures that made them feel like a potential criminal by virtue of what they experienced as both intrusive and opaque. Yet, racialised forms of surveillance extended beyond technological systems and disrupted participants' ability to establish everyday security through predictability and routines in their daily lives (cf.~\Cref{sec:hci-border-tech}). For example, participants' feelings of criminalisation and suspicion were reinforced through bureaucratic opacity, repeated document checks and the obligation to ``prove innocence''. Thus, practices of `othering' were co-produced by technological and administrative systems, while feelings of being `othered' transcended the border itself (cf.~\Cref{sec:memories-border}). In~\cite{Sociology:MytWalKha13}, the authors highlight how counter-terrorism measures in the UK such as extensive `stop and search' measures and the deployment of control orders has resulted in a division of security for some groups and insecurity for others. We found in our analysis that feelings of `othering' were produced through UK Government measures that afforded protection to some and not to `others'. 

\subsection{Implications for Design}\label{sec:implications}
Our interventions are grounded in a bounded empirical study (both in time and place). While this enables in-depth qualitative inquiry into individual and collective experiences, our design recommendations should be understood as theoretically informed provocations requiring further participatory engagements across different settings and with a diversity of people~\cite{10.1145/1124772.1124855}. Thus, following established practice in exploratory HCI research, our contribution lies in identifying what \emph{might} require redesign in light of our findings; thereby surfacing design tensions that merit systematic investigation rather than providing solutions ready for implementation.

\subsubsection{Extending Participatory Design: Before the Border}\label{sec:pre-border-design}

HCI scholarship has demonstrated the importance of participatory design with forced migrants, establishing methods for creating safe participation spaces~\cite{bustamante2019safe}, addressing cultural difference and power asymmetries~\cite{bustamante2018participatory} and co-designing technologies that account for literacy, health beliefs and privacy concerns~\cite{Noyman2017,talhouk2016syrian}. These works have primarily focused on post-arrival experiences: digital service access during resettlement~\cite{coles-kemp2019accessing,coles-kemp2018new}, healthcare system navigation~\cite{talhouk2016syrian} and accommodation planning~\cite{Noyman2017}; they share a temporal assumption: that design work begins after migration has occurred, once the border has been crossed.

The participants we engaged through our study shared how they had were both confused and anxious when they first encountered biometric technologies -- such as M1's unfamiliarity with iris scanning and R2's disorientation during fingerprinting (cf.~\Cref{sec:meeting-the-border}). This suggests that one avenue for future participatory design practice could be realised through pre-border interventions grounded in trauma-informed computing. Research in trauma-informed computing emphasises how considering traumatic stress reactions provides new insights into people's technology experiences, advocating for safety, trust and collaboration as core design principles~\cite{chen2022trauma,CSCW:RSANDW24}. We propose collaborative design interventions that draw on the experiences of migrants who have successfully navigated border control processes. For example, our findings highlight how caseworkers drew on their own experiences when supporting clients navigating screening processes (cf.~\Cref{sec:meeting-the-border}). Others have also shown how both formal and informal support structures such as caseworkers~\cite{simko2018} and language teachers~\cite{coles-kemp2019accessing,coles-kemp2018new} are critical to a successful resettlement.

These works, like our study, highlight the fluidity of the border. In a narrow sense, the border and the technologies encountered at the border are bounded by time and space; yet, the border and how it is experienced extends beyond the physical border itself. We therefore also consider a pre-border preparation infrastructure developed through collaborations between migrants who have successfully navigated screening processes, caseworkers whose own migration histories foster client trust and HCI researchers to potentially alleviate some of the uncertainty voiced by participants. While \Cref{sec:caseworkers} exemplifies the three types of work (explanatory, affective and procedural) performed by caseworkers at and after the border, pre-border materials could begin this work earlier. Multilingual instructional materials such as videos, simulations and pictorial guides that explain UK border screening procedures before arrival might help alleviate some of the fear voiced by several participants: that confusion or technical errors during border screening could jeopardise their asylum claims and risk their future in the UK (cf.~\Cref{sec:fear-tech-border}). These materials would not eliminate the anxiety of border crossing but could reduce the shock of encountering unfamiliar technologies at a moment of vulnerability, creating a continuum of support that begins before arrival and continues through the checkpoint encounter into resettlement. Yet, it is worth noting that support organisations, such as the charity at the heart of our study, are themselves interwoven in the border context and shaping how it is perceived and negotiated.

This temporal gap, where confusion and fear were experienced first and understanding only facilitated later through caseworker interpretations, suggests shifting organisational mediation forward rather than leaving it as post-hoc repair. The three types of work we identified in \Cref{sec:caseworkers} -- explanatory, affective and procedural -- thus become design suggestions derived from observing what happens when refugees and asylum seekers are excluded from design processes. Translating these suggestions into actual systems would benefit from collaborative approaches in which refugees help shape both training materials for officials and procedural protocols, drawing on their knowledge of what feels threatening or reassuring, yielding interventions across multiple dimensions~\cite{amnesty2024digitalborder,Bobeth2013}.

\subsubsection{Designing the Border}\label{sec:designing-border}

Our findings reveal a core tension. Border technologies are intended to process migrants efficiently, yet they often erode the agency of those being processed. Participants described feeling unable to ask questions, challenge decisions or understand what was happening to them and their data. For example, this is underpinned by R3's reflection that there was no-one to ask and no-one to offer reassurance when they first encountered border screening. Their experiences capture how the system created a sense of agency loss where R3, like others in their situation, had to comply and provide information without being able to comprehend or influence the procedures that would shape their future. However, as our findings show (cf.~\Cref{sec:caseworkers}) agency may be partly restored through organisational mediation, where sense-making occurs in he relationship between client, caseworker and infrastructure. Similar observations are made by the authors of~\cite{coles-kemp2018new} and~\cite{simko2018}. This shifts the design object for HCI from individual devices to an assemblage of people, artefacts and practices that together shape how technology is experienced. Further, trauma-informed HCI research, e.g~\cite{chen2022trauma}, points to how safety, trust and collaboration are not `extras' but preconditions for technology use among traumatised groups and individuals.

Within this assemblage, supporting agency begins with making technological processes more legible. Explanations that caseworkers already provide could be adapted into multilingual pictorial walk-throughs at immigration or border check-points that show what biometric devices collect, why data is needed and how each step progresses~\cite{bustamante2018participatory}. Trauma-informed computing principles, which centre on safety, trust and collaboration emphasise quieter scanner alerts, real-time visual displays of what devices detect and spatial layouts that avoid interrogation aesthetics, each of which can reduce fear among those who associate biometric submission with past experiences of violence~\cite{chen2022trauma}. Simple checklists, status updates and error prevention prompts can reduce cascading failures while signalling that users' experiences are taken seriously~\cite{Fisher2022}. The contrast between R1's experience with an official who showed a degree of empathy and the many who felt treated as potential criminals (cf.~\Cref{sec:dehumanising}) demonstrates how procedural conduct directly shaped how participants experienced the border.

Considering caseworkers as part of the border interface reframes border processing as a distributed sociotechnical arrangement that can both harm and support. Rather than technology alone determining migrants' experiences, those experiences might be better supported through the interpretive work that caseworkers perform under structural constraints. This brings to light a tension at the centre of our study: the same assemblage that harms through opacity and surveillance relies on intimate forms of carework to remain navigable~\cite{Fisher2022}. Holding on to this tension analytically allows HCI research to explore some forms of harm reduction and more inclusive design practices without assuming that technological interventions can resolve the deeper structural inequalities embedded in border regimes.

\subsection{Participatory Design Challenges}\label{sec:design-challenges}

Our proposed design interventions confront significant barriers in existing border control regimes. Resistance may come from border agencies viewing transparency as threatening to security functions, particularly given the ``presumption of guilt'' ideology identified by CW5 (cf.~\Cref{sec:across-border}). This ideological foundation, that migrants must prove legitimacy rather than be presumed legitimate, structures how border technologies are commissioned and deployed. Redesigning technological interfaces requires hardware, software and training investments that governments may resist when current systems are perceived as functional. The contrast uncovered in our study, between Ukrainian refugees receiving a ``red carpet welcome'' and Afghan or African asylum seekers facing ``protracted legal struggles'', demonstrates how border processing reflects perceived political hierarchies that technological redesign alone cannot address.

Tensions between control and care underpin these design interventions and challenge core immigration ideologies, requiring broader societal transformation rather than design solutions alone. There is a risk of co-optation: making technologies more `humane' may legitimise border surveillance itself, potentially reducing pressure for systemic changes. This connects to the tension we identified in \Cref{sec:designing-border}; the same assemblage that harms through opacity requires intimate carework to remain navigable. Designing `better' technologies within this assemblage risks stabilising it rather than challenging it. 

Participatory design in border contexts requires methodological adaptation. Legal precarity means traditional co-design approaches may be inappropriate: participants may feel unable to honestly critique systems determining their legal status~\cite{edri2020humanrights-migrationtech}. Trauma-informed approaches must thus ensure participation itself does not recreate power dynamics experienced during border processing~\cite{Voith2020}. Moreover, recognising that experiences and memories of border technology were shared by most participants in our study, interventions addressing collective rather than individual experiences are likely both more fruitful. Community-based participatory research frameworks emphasise survivor-centred approaches shifting power dynamics and enabling bi-directional learning~\cite{Jumarali2021,Bobeth2013}. Meaningful participatory design therefore requires long-term engagement supporting both immediate interface improvements and broader advocacy for systemic change.

Despite these constraints, the experiences shared by participants demonstrate that design choices matter. CW1's lingering anxiety decades after arrival -- ``even now, when I go through some of these screenings'' -- suggests that initial encounters structure everyday security long after legal status is secured. Thus, we put forward, with caution, these interventions not as comprehensive solutions but as incremental changes that may reshape elements of border security while broader systemic failures continue.

\section{Conclusion}\label{sec:conclusion}

We have shown how the initial encounter with border technologies in the UK reverberates far beyond the checkpoint itself and shapes asylum seekers' and refugees' ongoing sense of security, belonging and trust in institutions in their host country. Our findings reveal that automated screening systems frequently produce confusion, fear and feelings of dehumanisation, which are only partially mitigated by the presence of caseworkers. In line with HCI traditions of participatory and critical design, we call for future work that expands co-design to include caseworkers and migrants as co-creators of border technologies and support tools. Such work would benefit from trauma-informed computing, multilingual accessibility and survivor-centred participatory research practices to reimagine how border encounters could prioritise emotional responses and everyday feelings of security. More broadly, our study underscores the importance of centring lived experiences in debates on digital border infrastructures, pointing towards HCI's capacity to explore the intersections of everyday security, technology and carework in everyday life.

\paragraph{Limitations.} \emph{Single-site fieldwork}: Our observations and interviews were conducted with one London-based charity and are thus limited to this specific setting. 
\emph{Language mediation}: We had to rely on caseworkers for interpretation and translation which may have shaped how the voices of participants were translated and understood, which stories they shared, and, thus, captured in our work. We reflect on this in detail in~\Cref{sec:interviews-method}.
\emph{Temporal scope}: Participant observation at the charity lasted three months, which is a relatively short time-frame to enable immersion in the setting under study. This also means that our work does not capture all dynamics within the charity and between clients and caseworkers, but was bounded by the experiences of those who were present in the space at the time of the study. Longer participation might reveal subtle and evolving dynamics of resettlement and technological adaptation, over time. 
\emph{Positionality of the researcher}: As a migrant embedded in the field site, the researcher who conducted the observational work had her own experiences with UK border systems that may have shaped interactions with participants. Although reflexivity was built into the analysis, our positionalities shaped our interpretations of the data. This is, however, not uncommon in this type of work.

\section*{Acknowledgements}
This work would not exist without the many contributions from the participants at the centre, who trusted us with their stories and experiences. The research of Dutta was supported by the UKRI as part of the Centre for Doctoral Training in Cyber Security for the Everyday at Royal Holloway, University of London (EP/S021817/1).

\bibliographystyle{plain}
\bibliography{local.bib}

\appendix

\section{Interview Guides}\label{sec:interview-guides}

\subsection{Interview Guide: Asylum Seekers and Refugees}\label{sec:interview-immigrants}

\begin{enumerate}
	\item Experiences at the Border:
	\begin{itemize}
		\item Can you describe your experience when you arrived at the UK border? Were there any aspects that made you feel particularly uncomfortable or welcome?
		\item Which technologies did you encounter at the border?
		\item How did you feel about the technology used during your screening process, such as biometric scans or automated systems?
	\end{itemize}
	\item Perceptions of Privacy and Security:
	\begin{itemize}
		\item How do you feel about the collection and use of your personal and biometric data by border authorities? Do you have any concerns about your privacy?
		\item Do you feel that the border control processes are secure? Why or why not?
	\end{itemize}
	\item Impact of Border Technologies:
	\begin{itemize}
		\item In what ways, if any, did the use of technology at the border affect your perception of safety and fairness in the immigration process?
		\item Did you notice any differences in treatment based on your background or appearance? How did this impact your experience?
		\item How did such differences make you feel? More or less secure?
	\end{itemize}
	\item Ethical and Human Rights Considerations:
	\begin{itemize}
		\item Do you feel that your rights were respected throughout the border control process? Were you aware of any channels to report concerns or seek help?
		\item Did you feel your individual rights were respected at the border?
		\item How can that be better balanced with security measures?
	\end{itemize}
	\item Trust and Transparency:
	\begin{itemize}
		\item To what extent do you trust border authorities and the technologies they use? What factors influence this?
		\item Did you feel informed about how your data would be used and stored? What kind of information or assurances would help you feel more at ease?
	\end{itemize}
	\item Long-Term Implications:
	\begin{itemize}
		\item How has your experience at the border affected your view of the UK and its approach to immigration?
		\item What are your hopes for how future immigrants will be treated at UK borders?
	\end{itemize}
	\item Suggestions for Improvement:
	\begin{itemize}
		\item What changes would you suggest to make the border screening process more efficient?
	\end{itemize}
\end{enumerate}

\subsection{Interview Guide: Caseworkers}\label{sec:interview-caseworkers}

\begin{enumerate}
	\item Understanding Current Practices:
	\begin{itemize}
		\item Could you describe the current border screening technologies that your clients typically encounter when arriving in the UK?
		\item How do these technologies impact the experience of refugees and asylum seekers during their initial entry and subsequent interactions with border officials?
	\end{itemize}
	\item Challenges and Concerns:
	\begin{itemize}
		\item What are the most common concerns or challenges that refugees and asylum seekers face regarding these border screening technologies?
		\item Have you observed any specific issues related to privacy or data security in these technologies?
	\end{itemize}
	\item Support and Advocacy:
	\begin{itemize}
		\item How does [this charity] assist refugees in navigating and understanding these border screening processes?
		\item Are there any particular areas where you feel additional support or advocacy is needed to help refugees deal with these technologies?
	\end{itemize}
	\item Feedback and Experiences:
	\begin{itemize}
		\item Could you share any feedback or experiences from refugees that highlight the impact (positive or negative) of border screening technologies on their sense of safety and well-being?
		\item Have any of your clients expressed specific frustrations or anxieties about the use of biometric or AI-based screening tools?
	\end{itemize}
	\item Ethical and Responsible AI:
	\begin{itemize}
		\item What are your thoughts on the ethical implications of using AI-based border screening technologies in immigration control?
	\end{itemize}
	\item Future Improvements:
	\begin{itemize}
		\item In your opinion, what improvements or changes to border screening technologies would most benefit refugees and asylum seekers?
	\end{itemize}
\end{enumerate}

\onecolumn

\section{Example Reflexive Coding Tables}\label{sec:reflexive-coding-table}

\begin{figure}[h!]
	\includegraphics[keepaspectratio,height=.48\textheight]{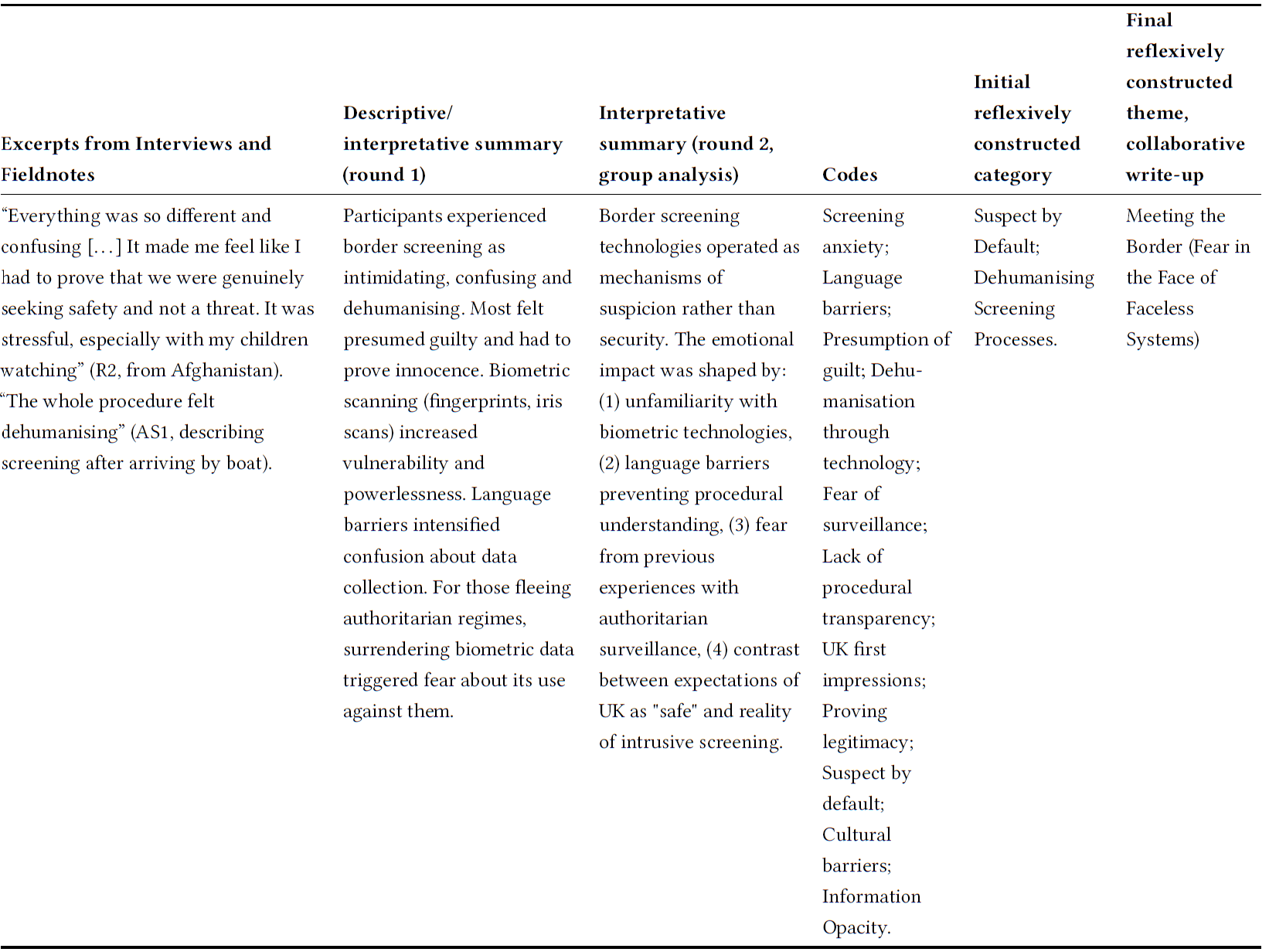}
\end{figure}

\begin{figure}[h!]
	\includegraphics[keepaspectratio,height=.41\textheight]{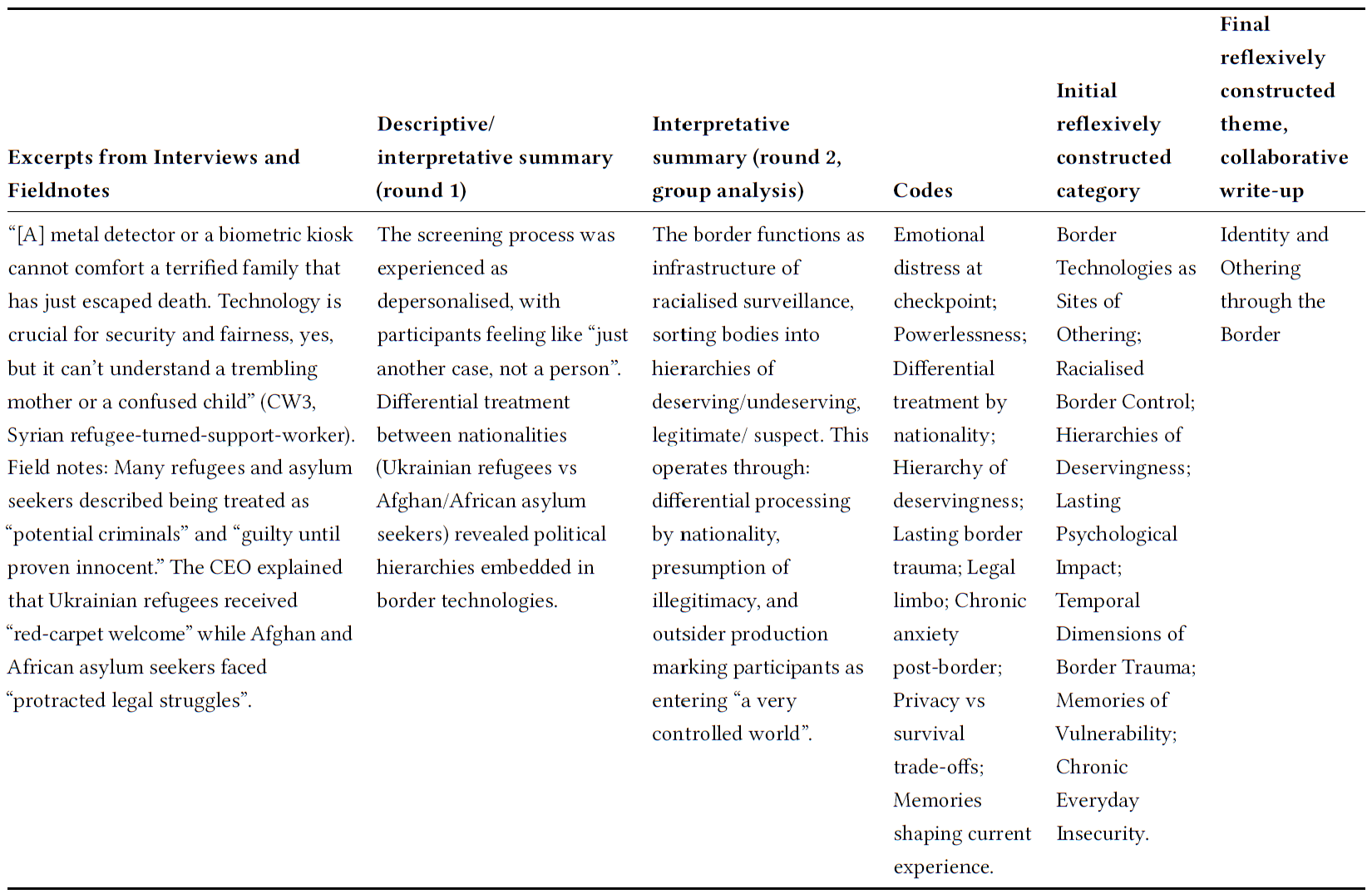}
\end{figure}

\begin{figure}[h!]
	\includegraphics[keepaspectratio,height=.43\textheight]{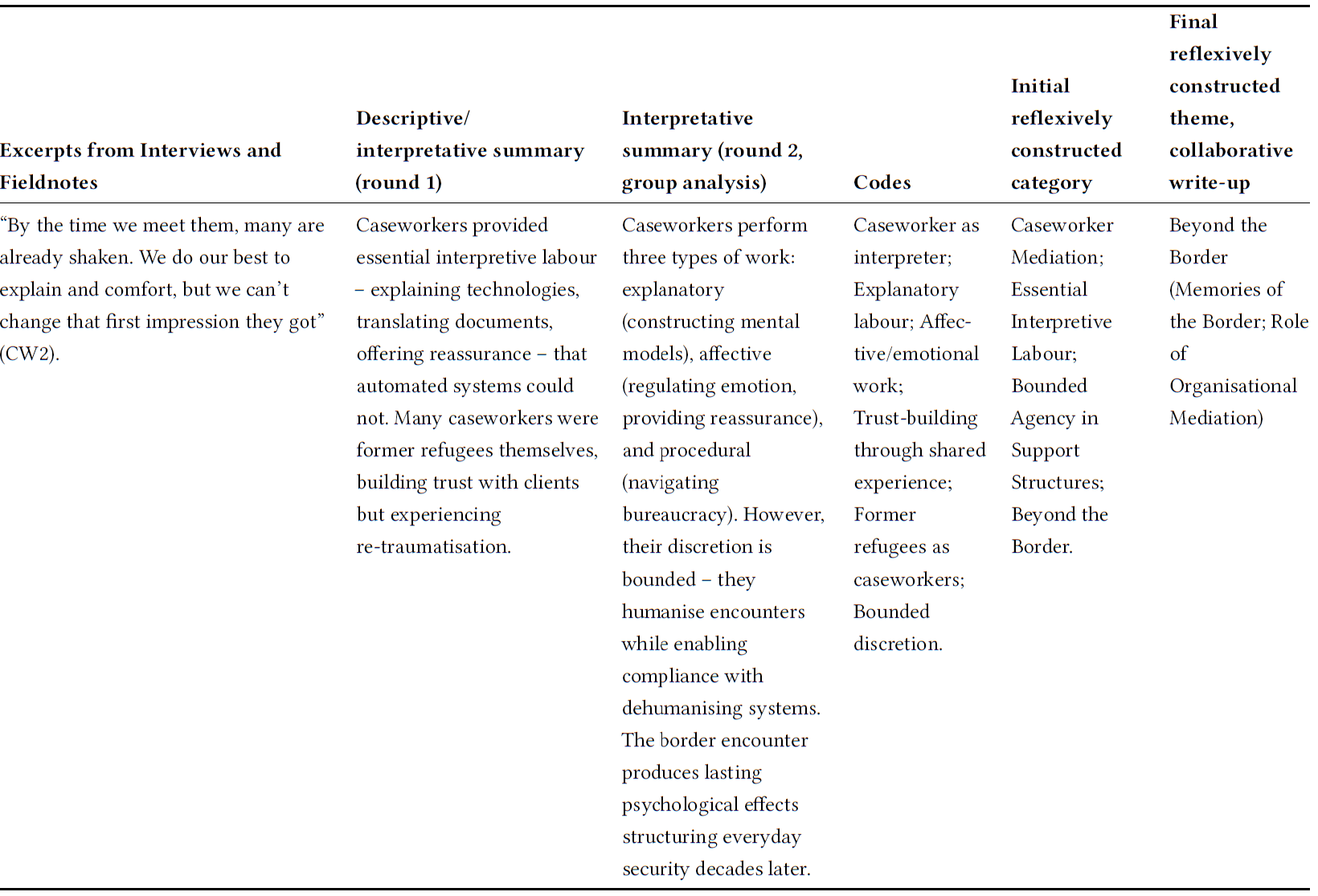}
\end{figure}

\end{document}